\documentclass[a4paper,11pt]{article}
\pdfoutput=1 

\usepackage{jcappub} 

\usepackage[normalem]{ulem}

\usepackage[T1]{fontenc} 
 \usepackage{xcolor}
 \newcounter{attnctr} 

\title{\boldmath Nonparametric test of consistency between cosmological models and multiband CMB measurements}

\author[a]{Amir Aghamousa,}
\author[b]{and Arman Shafieloo}

\affiliation[a]{Asia Pacific Center for Theoretical Physics,\\Pohang, Gyeongbuk 790-784, Korea}
\affiliation[b]{Korea Astronomy and Space Science Institute,\\Daejeon, 305-348 Korea}

\emailAdd{amir@apctp.org}
\emailAdd{shafieloo@kasi.re.kr}

\abstract{We present a novel approach to test the consistency of the
cosmological models with multiband CMB data using a nonparametric
approach. In our analysis we calibrate the REACT (Risk Estimation and Adaptation after Coordinate Transformation) confidence levels associated with distances in function space (\textit{confidence distances}) based on the Monte Carlo simulations in order to test the consistency of an assumed cosmological model with observation. To show the applicability of our algorithm, we confront Planck 2013 temperature
data with concordance model of cosmology considering two different
Planck spectra combination. In order to have an accurate quantitative statistical measure to compare
between the data and the theoretical expectations, we calibrate REACT confidence distances 
and perform a bias control using many realizations of the data. Our results in this work using Planck 2013 temperature data  
put the best fit $\Lambda$CDM model at $95\% (\sim 2\sigma)$ confidence distance from the center of the
nonparametric confidence set while repeating the analysis excluding the Planck $217 \times 217$
GHz spectrum data, the best fit $\Lambda$CDM model shifts to $70\% (\sim 1\sigma)$ confidence distance. 
The most prominent features in the
data deviating from the best fit $\Lambda$CDM model seems to be at low
multipoles $ 18 < \ell < 26$ at greater than $2\sigma$,
$\ell \sim 750$ at $\sim1$ to $2
\sigma$ and $\ell \sim 1800$ at greater than $2\sigma$ level. Excluding
the $217\times217$ GHz spectrum the feature at $\ell \sim 1800$ becomes
substantially less significance at $\sim1$ to $2 \sigma$ confidence level. Results of our
analysis based on the new approach we propose in this work are in
agreement with other analysis done using alternative methods.}

\begin{document}
\maketitle
\flushbottom

\section{Introduction}
Cosmic microwave background (CMB) observations currently provide the
most powerful probe of the history and constituents of the Universe.
The analyses of the Wilkinson Microwave Anisotropy Probe (WMAP)
\citep{Bennett2013, Hinshaw2013} and Planck \citep{Planck2013I} as the
two full sky surveys of CMB approve the overall validity of the
standard $\Lambda$CDM model with six cosmological parameters.
In spite of the agreement on the cosmological framework, Planck has
derived somehow different cosmological parameters than what were
obtained by WMAP in combination with other ground-based CMB surveys
such as the Atacama Cosmology Telescope (ACT, \cite{Seievers2013ACT})
and South Pole Telescope (SPT, \cite{Story2013SPT}). There have been
so far few investigations to address this ambiguity.
\cite{Planck2013XV, Planck2013XVI, Planck2013XXXI} and
\cite{Kovacs2013} discussed that the WMAP angular power spectrum seems
to have a higher amplitude with respect to Planck's angular power
spectrum at some range of multipoles. \cite{Hazra2014b} reported
consistency of the angular power spectrum data from WMAP and Planck
provided an overall amplitude shift and a tension at 3$\sigma$ level
with fixed amplitudes. In another attempt \cite{Hazra2014a} showed
that the concordance model of cosmology is consistent to Planck data
only at 2 to 3$\sigma$ confidence level. Recently \cite{Larson2014}
performed a comprehensive consistency test of WMAP 9-year data and the
Planck 2013 data. They found $\sim2.5\%$ difference in spectra ($\ell
\gtrsim 100$) is significant at the $3$-$5\sigma$ level.

In this paper we test the consistency of the concordance model of
cosmology with Planck 2013 data by implementing and calibrating a
nonparametric approach known as REACT (Risk Estimation and Adaptation after Coordinate Transformation). Using REACT as a
nonparametric and model-independent methodology, \cite{Aghamousa2012}
checked the consistency of WMAP 1, 3, 5, 7-year data with the standard
model. This methodology provides a confidence metric that measures the
consistency of the angular power spectrum data and the expectations of
a given model. The method was also recently used in
\citep{Aghamousa2014b} and results indicated that the concordance
model of cosmology is consistent with Planck 2013 data at the level of
$36\% (\sim 0.5\sigma)$. While REACT is a strong statistical framework
to test consistency of the models and the data, it is very much
conservative in its internal measures. In this paper we calibrate
REACT confidence distances by performing many Monte Carlo simulations to have a more
precise measurement of the consistency between the Planck 2013 data
and the concordance model. We will show that after calibration, our
analysis indicates that the consistency between the standard model and
Planck 2013 data is only about 2$\sigma$ confidence level.
Since there have been various discussions about possible systematic in
the Planck 217$\times$217 GHz power
spectrum~\cite{Planck2013XXII, Spergel2013}, we do also perform the
analysis excluding this spectrum.

In what follows, Section~\ref{subsec:data} describes the two sets of
weighted-average angular power spectrum data that we used in this
work. We discuss nonparametric methodology we used in this paper in Section~\ref{sec:methodology} which starts with a review of estimating nonparametric fit and confidence set (Subsection~\ref{subsec:nonparametricfit}) and validating cosmological models (Subsection~\ref{subsec:validating}) then continues with description of our extensions, calibrating the confidence distances (Subsection~\ref{subsec:calibrating}) and bias control (Subsection~\ref{subsec:bias}).
In Section~\ref{sec:results} we present the
results and finally we conclude this work in Section~\ref{sec:conclusions}.

\section{Data}~\label{subsec:data}
The Planck satellite measured the CMB temperature anisotropies in a
wide range of channels from 30 to 353 GHz and provides the temperature
angular power spectrum in the multipole range of $2\leq \ell \leq
2500$ \citep{Planck2013I}. At the multipoles $l<50$, the angular power
spectrum is derived by using all frequency channels from 30 to 353 GHz
to remove the Galactic foreground. For multipoles $\ell \geq50$, the
angular power spectra in the frequency range from 100 to 217 GHz are
contaminated significantly by extragalactic foregrounds. To consider
these foregrounds appropriately Planck likelihood code implements a
foreground model including set of nuisance parameters to calculate the
likelihood of a cosmological model to the data \citep{Planck2013XV, Planck2013XVI}.

To provide a single CMB angular power spectrum data, as we need to use
in our analysis, we follow the same procedure as in
\citep{Aghamousa2014b}.
Using the nuisance parameters associated with the Planck best fit
$\Lambda$CDM \citep{Planck2013XVI, Planck2013XV} we obtain the
background angular power spectra data in the frequency range 100 to
217 GHz. Then we calculate the weighted-average of angular power
spectra to inverse of their corresponding variance between the
contributing frequency channels in each multipole $\ell$. The
weighted-average angular power spectrum, $\bar{{\cal D}_{\ell}}$ would
be

\begin{eqnarray}\label{equ:weight-average1}
\bar{{\cal D}_{\ell}} &=& \frac{\sum_{ch}w^{ch}_{\ell} {\cal
D}^{ch}_{\ell}}{\sum_{ch}w^{ch}_{\ell}}, \\
 w^{ch}_{\ell}&=&(\sigma^{ch}_{\ell})^{-2} \nonumber
\end{eqnarray}
where ${\cal D}^{ch}_{\ell}=\ell (\ell+1)C_{\ell}/2\pi$ refers to
temperature angular power spectrum of associated frequency channel in
each $\ell$.
The weight term $w^{ch}_{\ell}$ equals to inverse variance,
$(\sigma^{ch}_{\ell})^{-2}$, in each $\ell$ and $ch$ runs for
contributing frequency channels.
The corresponding covariance matrix takes the form
\begin{eqnarray}\label{equ:weight-covariance}
\text{Cov}(\bar{{\cal D}_{\ell}},\bar{{\cal D}_{\ell'}}) &=&
\frac{1}{\sum_{ch}w^{ch} \sum_{ch'}w^{ch'}} \quad
\text{Cov}(\sum_{ch}w^{ch}_{\ell} {\cal
D}^{ch}_{\ell},\sum_{ch'}w^{ch'}_{\ell'} {\cal D}^{ch'}_{\ell'})
\nonumber \\
&=& \frac{1}{\sum_{ch}w^{ch} \sum_{ch'}w^{ch'}} \sum_{ch, ch'}
w^{ch}_{\ell} w^{ch'}_{\ell'} \text{Cov}( {\cal D}^{ch}_{\ell}, {\cal
D}^{ch'}_{\ell'})
\end{eqnarray}
where $ch$ and $ch'$ run for the contributing frequency channels and
$\text{Cov}( {\cal D}^{ch}_{\ell}, {\cal D}^{ch'}_{\ell'})$ is the
covariance of frequency channels $ch$ and $ch'$ at different
multipoles $\ell$, $\ell'$ (provided by Planck likelihood code).

\begin{figure}
   \includegraphics[width=\textwidth]{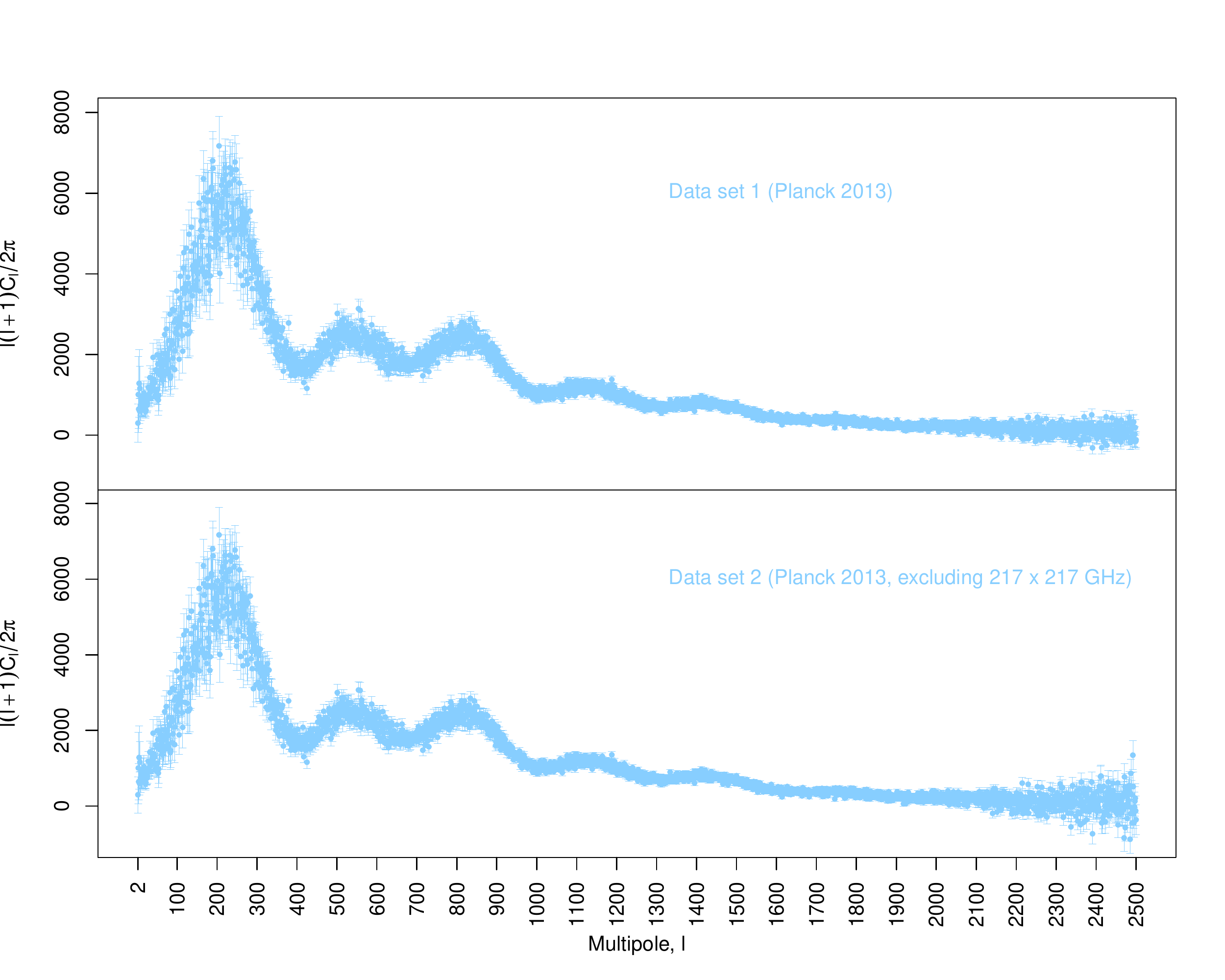}
  \caption{\label{fig:dataset1} The weighted-average Planck 2013
temperature angular power spectrum data. Data set 1, is the
weighted-average of the background CMB angular power spectra from
combination of $100 \times 100$ GHz, $143\times143$ GHz,
$217\times217$ GHz and $143\times217$ GHz spectra at ($\ell \ge50$).
Data set 2, is similar to data set 1, except that the angular power
spectrum of $217\times217$ GHz is excluded. The angular power spectrum
at $\ell<50$ in both data sets is given by Planck team and obtained by
using the Planck frequency channels from 30 to 353 GHz
\citep{Planck2013XV}.}
 \end{figure}

Using above procedure we obtain two sets of wighted-angular power
spectrum data and corresponding covariance matrices from Planck 2013
data.
The angular power spectrum \textit{data set 1} is made by using the all frequency channels
at multipoles $\ell \geq50$ (listed in Table~\ref{tab:cross-spectra}).
In addition we calculate the angular power spectrum \textit{data set
2}, similar to the \textit{data set 1} but excluding the Planck
$217\times217$ GHz angular power spectrum.
The Figure~\ref{fig:dataset1} depicts both data sets. We see at
multipoles greater than $\ell \sim 2000$ the noise level of data set
2 is higher than data set 1 which is due to excluding the the
$217\times217$ GHz angular power spectrum.

\begin{table}[!htb]
\begin{center}
\vspace{6pt}
\begin{tabular}{| l | c |}
\hline
Spectrum  &  Multipole range\\
\hline
$100\times100$  &  50-1200   \\
$143\times143$  &  50-2000   \\
$217\times217$  &  500-2500  \\
$143\times217$  &  500-2500  \\
\hline
\end{tabular}
\end{center}\caption{~\label{tab:cross-spectra} The multipole ranges
of the spectra and the cross-spectra provided by Planck 2013 data for
$\ell \ge 50$ \citep{Planck2013XV}.}
\end{table}

\section{Methodology}~\label{sec:methodology}
The REACT is a nonparametric function estimation and was initially proposed by \citep{Beran1996, BD1998,
Beran2000, Beran2000b} and then modified and used in CMB data analysis by \citep{GMN+2004, BSM+2007, Aghamousa2012, Aghamousa2014,
Aghamousa2014b}.

The function estimation or regression methods estimate the unknown function $f$, related to given data set, by using the estimate $\widehat{f}$. 
The nonparametric function estimation methodologies do not assume a specific functional form for $f$ and attempt to estimate the function with some mild regularity assumptions such as smoothness or membership of $f$ to a function space, etc. In REACT formalism an estimate $\widehat{f}$ of unknown function $f$ is obtained by balancing 
bias and variance of $\widehat{f}$ through optimal smoothing which is achieved by minimizing the risk function. This nonparametric methodology also constructs a confidence set around the $\widehat{f}$ which can be employed in many aspects of inference.

In this section we first present a review of this methodology (Subsections~\ref{subsec:nonparametricfit} and \ref{subsec:validating}) which is mainly based on \citep{GMN+2004, BSM+2007, Aghamousa2012}. Our calibration of confidence distances based on the Monte Carlo simulations in order to test the consistency of an assumed cosmological model with observation, is presented in Subsection~\ref{subsec:calibrating}. In addition we describe (Subsection~\ref{subsec:bias}) a way of bias estimation of $\widehat{f}$ based on the Monte Carlo simulations to demonstrate the consistency in multipole space.

\subsection{The nonparametric fit and confidence set}~\label{subsec:nonparametricfit}
The CMB angular power spectrum data is given in the form of 
 \begin{equation}
  Y_\ell = {\cal D}_{\ell} + \epsilon_\ell
 \end{equation}
where $Y_\ell$ repersents the $N$ observed data points over multipoles equal to true but unknown angular power spectrum ${\cal D}_{\ell} \equiv f(x_{\ell})$ contaminated with the noise, $\epsilon_l$, which is assumed to have normal distribution with mean-0 and known covariance matrix $\Sigma$.

The main assumption in REACT formalism is that the unknown function $f$, belongs to $L_2$ function space, 
such as it can be expanded in a complete orthonormal basis $\{ \phi_j(x) \}$,
$$f(x) = \sum_{j=0}^\infty \beta_j \phi_j(x).$$
A proven useful basis for the CMB angular power spectrum estimation is the cosine basis defined over $0 \le x \le 1$:
 \begin{equation}
  \label{eq:cosinebasis}
  \phi_j(x) = \begin{cases} 1 \;\;\;\;\;\;\;\;\;\;\;\;\;\;\;\;\;\; (j=0) \\ \sqrt{2} \cos(j\pi x) \; (j = 1, 2, \ldots) \end{cases}.
 \end{equation}

Assuming $f$ is sufficiently smooth and considering $N$ data points, we can take 
$$f(x) \approx \sum_{j=0}^{N-1} \beta_j \phi_j(x),$$
 and estimate it as 
 \begin{equation}
  \label{eq:fhat}
  \widehat{f}(x) = \sum_{j=0}^{N-1} \widehat{\beta}_j \phi_j(x).
 \end{equation}

The angular power spectrum ${\cal D}_{\ell} \equiv f(x_{\ell})$ would be estimated as $\widehat{{\cal D}}_{\ell} \equiv \widehat{f}(x_{\ell})$  through coefficient estimates $\widehat{\beta}_j$ which are estimated as
 \begin{equation}
  \label{eq:betahat}
  \widehat{\beta}_j = \lambda_j Z_j,
 \end{equation}
 where
 \begin{equation}
  \label{eq:Z}
  Z_j = {1 \over N} \sum_{i=0}^{N-1} Y_i \phi_j(x_i) = {( U^T Y )_j \over \sqrt{N}}.
 \end{equation}
Here, $U$ is the orthonormal matrix defined as $U_{ij} = \phi_j(x_i) / \sqrt{N}$ and $Y \equiv (Y_0,\ldots,Y_{N-1})^T$

 Finding the coefficient estimates $\widehat{\beta} \equiv ( \widehat{\beta}_0, \ldots, \widehat{\beta}_{N-1} )^T$,
 and hence the $\widehat{f}(x)$, could be reduced to determining the \emph{shrinkage parameters} $\lambda_j$. The smoothness assumption of $f$ implies a fast decay of the true coefficients $\beta_j$ with $j$. This suggests a monotonically decreasing constraint on shrinkage parameters
 \begin{equation}
  \label{eq:monotone}
  1 \ge \lambda_0 \ge \lambda_1 \ge \ldots \ge \lambda_{N-1} \ge 0 \;\; \mbox{(Monotone shrinkage)}.
 \end{equation}

Using shrinkage parameters one can define the \emph{effective degree of freedom} (EDoF) as
 \begin{equation}
  \label{eq:edof}
  \mbox{EDoF}(\lambda) = \sum_{j=0}^{N-1} \lambda_j.
 \end{equation}
Effective degree of freedom can be used to characterize the nonparametric function estimation $\widehat{f}(x)$ and interpreted as the degree of smoothness.

In this formalism the difference between the true unknown function $f(x)$ and its estimator $\widehat{f}(x)$ is measured via 
the inverse-noise-weighted \emph{loss function} $L(\lambda)$,
 \begin{equation}
  \label{eq:loss}
  L(\lambda) = \int_0^1 \left( { f(x) - \widehat{f}(x) \over \sigma(x) } \right)^2 dx.
 \end{equation}
where $\sigma^2(x)$ is the variance of the data $Y$ and $\lambda \equiv (\lambda_0, \ldots, \lambda_{N-1})^T$ is the vector of shrinkage parameters.

Optimal smoothing is obtained by balancing the bias of the function estimator $\widehat{f}(x)$ with its variance,
by minimizing the risk function $R(\lambda)$, which is the expected value of loss function $L(\lambda)$,
 $$
  R(\lambda) =   \int_0^1 \left( { f(x) - \mathbb{E}\left( \widehat{f}(x) \right) \over \sigma(x) } \right)^2 dx
               + \int_0^1 \mathbb{E}\left[ \left( { \widehat{f}(x) - \mathbb{E}\left( \widehat{f}(x) \right) \over \sigma(x) } \right)^2 \right] dx.
 $$
Where the two terms are the integrated squared bias and the integrated variance of function estimator $\widehat{f}(x)$, both weighed by $1/\sigma^2(x)$.

Since the risk function $R(\lambda)$ depends on the unknown function $f$, it needs to be estimated through an estimator. 
A particular Stein's Unbiased Risk Estimator (SURE) \cite{Stein1981} of this risk take the following form

 \begin{equation}
  \label{eq:riskhat}
  \widehat{R}(\lambda) = Z^T \bar{D} W \bar{D} Z + \mbox{tr}( D W D B ) - \mbox{tr}( \bar{D} W \bar{D} B )
 \end{equation}
 where
 $D \equiv \mbox{diag}(\lambda_0, \ldots, \lambda_{N-1})$, $\bar{D} = I_N - D$,
  $Z \equiv (Z_0,\ldots,Z_{N-1})^T$, $B = U^T \Sigma U / N$ is the covariance of $Z$, and $I_N$ is the $N \times N$ identity matrix.
Using approximate expansion $(1 / \sigma^2(x)) \approx \sum_{j=0}^{N-1} w_j \phi_j(x)$,
the weight matrix $W$ would be defined as
 \begin{equation}
  \label{eq:W}
  W_{jk} = \sum_l \Delta_{jkl} w_l,
 \end{equation}
where,
 \begin{displaymath}
  \Delta_{jkl} = \int_{0}^{1} \phi_j(x) \phi_k(x) \phi_l(x) \hspace{1 mm} dx
               = \left\{
                  \begin{array}{ll}
                   1,                                                                        & \mbox{if }\#\{j,k,l=0\}=3, \\
                   0,                                                                        & \mbox{if }\#\{j,k,l=0\}=2, \\
                   \delta_{jk}\delta_{0l} + \delta_{jl}\delta_{0k} + \delta_{kl}\delta_{0j}, & \mbox{if }\#\{j,k,l=0\}=1, \\
                   \frac{1}{\sqrt{2}} (\delta_{l,j+k}+\delta_{l,|j-k|})                      & \mbox{if }\#\{j,k,l=0\}=0. \\
                  \end{array}
                 \right.
 \end{displaymath}
for the cosine basis (Eq.\ \ref{eq:cosinebasis}).

We have already mentioned the optimal nonparametric fit obtained by minimizing risk $\widehat{R}(\lambda)$ by $\widehat{\lambda}$ and through Equations~\ref{eq:fhat} \& \ref{eq:betahat}.
The estimated fit is the nonparametric fit without any restriction on its EDoF which is called the \textit{full-freedom fit}.
This full-freedom fit in CMB data analysis
can be quite wiggly in some multipoles where the noise levels in the
data is high. While this full-freedom fit is a reasonable reconstruction (in a
sense that it captures the essential trends in the data), most
cosmological models predicts smooth forms of the angular power
spectrum. 
Alternatively, imposing additional constraint of the form
 \begin{equation}
  \label{e:edof_q}
  \sum_{i=0}^{N-1} \lambda_i  = q,
 \end{equation}
where $q$ constrains the EDoF of the fit, leads to the smoother \textit{restricted-freedom fit}.
By gradually reducing the value of EDoF (relative to the full-freedom fit) we can obtain an acceptably smooth fit. 

Conventional curve estimation methods provide a confidence band around
the best fit that indicates the uncertainty in the reconstruction.
Instead, this nonparametric methodology quantifies the uncertainty
surrounding the nonparametric fit in the form of a high-dimensional
$(1 - \alpha)$ confidence set (\textit{confidence ball}) at a pre-specified $0 \le (1 - \alpha) \le 1$ confidence level. 
By definition the confidence ball asymptotically contains
the true model with probability equal to associated confidence level.
This confidence set for coefficient vector $\beta$ is centered at $\widehat{\beta}$ and is defined as
 \begin{equation}
  \label{eq:csetD}
  \mathcal{S}_{N,\alpha} = \left\{ \beta: (\beta-\widehat{\beta})^T W (\beta-\widehat{\beta}) \le {r}^2_\alpha \right\},
 \end{equation}
 where the \emph{confidence radius} ${r}_\alpha$ is calculated by
 \begin{equation}
  \label{eq:cradius}
  {r}^2_\alpha = {\widehat{\tau} z_\alpha \over \sqrt{N}} + \widehat{R}(\widehat{\lambda}).
 \end{equation}
 Here $z_\alpha$ is upper $\alpha$ quantile of standard normal distribution, and
 \begin{equation}
  \label{eq:csettau}
  \widehat{\tau}^2 / N = 2 \mbox{tr}(ABAB) + Z^T Q Z - \mbox{tr}(QB),
 \end{equation}
 where $Q = 4 ( ABA + WDBDW - 2 ABDW )$ and $A = DW + WD - W$.

On the other hand, loss function (Equation~\ref{eq:loss}) can be written approximately for $N$ data points as
 \begin{eqnarray}
  \label{eq:loss2}
  L(\lambda) & \approx &   {1 \over N} \sum_{\ell=\ell_{min}}^{\ell_{max}} \left( f(x_{\ell}) - \widehat{f}(x_{\ell}) \over \sigma(x_{\ell}) \right)^2 \\
						 &\approx& (\beta-\widehat{\beta})^T W (\beta-\widehat{\beta})
 \end{eqnarray}
Consequently the high-dimensional confidence set around the $\widehat{\beta}$ (Equation~\ref{eq:csetD}) can be interpreted as a 
confidence ball in a high-dimensional function space with the center of nonparametric fit $\widehat{f}$ weighed by $1/\sigma^2(x_{\ell})$,
with the same confidence radius ${r}_\alpha$ (Equation~\ref{eq:cradius}) to $(1 - \alpha)$ pre-specified confidence level.
Apart from the main application of confidence ball which is determining the uncertainties on the nonparametric fit and associated features, we can employ this for validating different cosmological models in a model-independent way. We will discuss this in the next subsection.

\subsection{Validating cosmological models}~\label{subsec:validating}
Basically the high-dimensional confidence ball, centered in the nonparametric fit, determines the uncertainties around the fit.
Furthermore we can use this concept for validating
different cosmological models against the data model-independently.
In this regard we measure the distance $d$ of the point associated to a
predicted angular power spectrum of a specific cosmological model to
the center of the confidence ball. 
Using Equation~\ref{eq:loss2} the distance, $d$, would be
 \begin{equation}
  \label{eq:distance}
  d = \sqrt{ {1 \over N} \sum_{\ell=\ell_{min}}^{\ell_{max}} \left( {\cal D}^{model}_{\ell} - \widehat{{\cal D}}_{\ell} \over \sigma_{\ell} \right)^2  }.
 \end{equation}
where $\widehat{{\cal D}}_{\ell}$ and ${\cal D}^{model}_{\ell}$ are
nonparametric and predicted angular power spectrum in each $\ell$
respectively.
The $\sigma^{2}_{\ell}$ is the variance of angular power spectrum data
in each $\ell$ and $N$ is the total number of multipoles.
This distance can be associated to the confidence level that a
proposed angular power spectrum can be rejected as a candidate for the
true model (\textit{confidence distance}).
For this purpose if we rearrange the Equation~\ref{eq:cradius}, the upper $\alpha$ quantile of standard normal distribution $z_\alpha (d)$, associated with distance $d$ would be
 \begin{equation}
  \label{eq:distance_alpha}
z_\alpha (d) = \frac{\sqrt{N}}{\widehat{\tau}} \left( d^2-\widehat{R}(\widehat{\lambda}) \right)
 \end{equation}
By definition the cumulative distribution function of the standard normal distribution of $z_\alpha (d)$ represents confidence distance (associated confidence level with particular distance $d$).
In other words through this procedure we can measure the
consistency of a specific cosmological model with the data in a
model-independent way.

\subsection{Calibrating confidence distances}~\label{subsec:calibrating}
Although the original formalism in REACT provides an elegant procedure to
assign a confidence level to a particular distance from the center of
nonparametric confidence set. However, in this procedure the
calculated confidence level is highly dependent on the minimum risk
value of the nonparametric fit  (Equation~\ref{eq:distance_alpha}) which is based on a single data
realization and REACT is usually very conservative in estimation of
inconsistencies between data sets and models. To address this issue
and in order to have a more accurate estimation of the level of
consistency (or inconsistency) between the concordance model of
cosmology and Planck 2013 data we perform a calibration based on many
simulated realizations of the angular power spectrum data, ${\cal
D}^{sim}_{\ell}$.
Since the main goal is to check the consistency of the Planck best fit
$\Lambda$CDM model with data, we assume the Planck best fit
$\Lambda$CDM as the true model and add the Gaussian noise according to
the calculated covariance matrix in
Equation~\ref{equ:weight-covariance}. Next we estimate the
full-freedom fit, $\widehat{{\cal D}}^{sim}_{\ell}$, of each simulated
angular power spectrum data.

Using Equation~\ref{eq:distance}, the distance of the Planck best fit
$\Lambda$CDM (true model) and the full-freedom fit, $\widehat{{\cal
D}}^{sim}_{\ell}$, of each simulated angular power spectrum data, can
be calculated. The \textit{empirical} Cumulative Distribution Function
(CDF) of these distances can be used as the calibrated confidence level of distances (\textit{calibrated
confidence distances}). Statistically, the obtained CDF
indicates the probability of existence of true model in distance of
shorter or equal to a specific distance in function space. In fact these calibrated confidence distances
present the (calibrated) confidence levels associated to distances from nonparametric fit in function space. 
We demonstrate these results in Section~\ref{sec:results}.

It is worth noting that while the nonparametric fit obtained by minimizing the risk estimator $\widehat{R}(\lambda)$ depends on the full covariance matrix of data $\Sigma$ (Equation~\ref{eq:riskhat}), the distance $d$ (Equation~\ref{eq:distance}) which is derived from the loss function uses only the diagonal terms of $\Sigma$.
This would not affect the reliability of the final results since we have used Monte Carlo simulations and treated the real and simulated data in a same way.
However, the method may become more sensitive to possible features in the angular power spectrum if the full covariance matrix in defining distances is employed.

\subsection{Bias control}~\label{subsec:bias}
In Subsection~\ref{subsec:nonparametricfit} we see the nonparametric fit is estimated by minimizing the risk
function that results to an optimally smooth reconstruction. In
principle, too much of smoothing leads to a fit with high bias and
low variance, and too little smoothing yields a fit with low bias and
high variance. Minimal risk (optimal smoothing) therefore can be
thought then as a reconstruction with a balance between the bias and
the variance. Therefore the nonparametric fit in this formalism has an
optimal smoothing with the cost of some bias.

Using the simulations we mentioned earlier, we are able to estimate
the bias and make our reconstructions more accurate. In this procedure
we use the distribution of the full-freedom fits, $\widehat{{\cal
D}}^{sim}_{\ell}$, in each $\ell$ obtained from our simulated angular
power spectrum data to find out the form of the bias. The median of
the distribution of $\widehat{{\cal D}}^{sim}_{\ell}$ in each $\ell$
represents an estimated angular power spectrum, $\widehat{{\cal D}}^{med}_{\ell}$, which tends to show true
angular power spectrum, ${\cal D}^{true}_{\ell}$. Therefore the bias
can be estimated as
 \begin{equation}
  \label{eq:bias}
  \text{bias}(\ell) = \widehat{{\cal D}}^{med}_{\ell} - {\cal D}^{true}_{\ell}
 \end{equation}
Then the calibrated unbiased nonparametric fit, $\widehat{{\cal
D}}^{unb}_{\ell}$, is obtained by subtracting the bias at each $\ell$
from the nonparametric fit, $\widehat{{\cal D}}_{\ell}$,
 \begin{equation}
  \label{eq:unbias}
  \widehat{{\cal D}}^{unb}_{\ell} = \widehat{{\cal D}}_{\ell} - \text{bias}(\ell)
 \end{equation}
 
The residual of Planck best fit $\Lambda$CDM and unbiased nonparametric fit can be calculated as
 \begin{equation}
  \label{eq:residual}
  \text{Res}(\ell) =  {\cal D}^{\Lambda CDM}_{\ell} - \widehat{{\cal D}}^{unb}_{\ell}
 \end{equation}
where $\text{Res}(\ell)$ is the residual value at each $\ell$ and the ${\cal D}^{\Lambda CDM}_{\ell}$ is the Planck best fit $\Lambda$CDM.
Since we assume the best fit $\Lambda$CDM as the true model in
simulation procedure (Subsection~\ref{subsec:calibrating}), then ${\cal
D}^{true}_{\ell} = {\cal D}^{\Lambda CDM}_{\ell}$. Therefore from
Equations~\ref{eq:bias}, \ref{eq:unbias} and \ref{eq:residual}, the
residual equation would be simplified to
 \begin{equation}
  \label{eq:residual2}
  \text{Res}(\ell) = \widehat{{\cal D}}^{med}_{\ell} - \widehat{{\cal D}}_{\ell}
 \end{equation}
The residual, $\text{Res}(\ell)$, represents the deviation of the Planck best
fit $\Lambda$CDM model from Planck 2013 data in different multipole
ranges estimated by the unbiased nonparametric reconstruction,
$\widehat{{\cal D}}^{unb}_{\ell}$.

In addition, the uncertainties of unbiased nonparametric fit, $\widehat{{\cal D}}^{unb}_{\ell}$ can be estimated by simulations.
Using the distribution of the full-freedom fits, $\widehat{{\cal D}}^{sim}_{\ell}$, in each $\ell$, we can estimate the corresponding $1$ and $2 \sigma$ error bars around $\widehat{{\cal D}}^{med}_{\ell}$. These uncertainties would be propagated identically to $\widehat{{\cal D}}^{unb}_{\ell}$, in each $\ell$, through Equations~\ref{eq:bias} \& \ref{eq:unbias}. They can be used to quantify the deviation of the Planck best fit $\Lambda$CDM model from the unbiased nonparametric reconstruction in terms of confidence levels.

\section{Results}~\label{sec:results}
\begin{figure}
   \includegraphics[width=\textwidth]{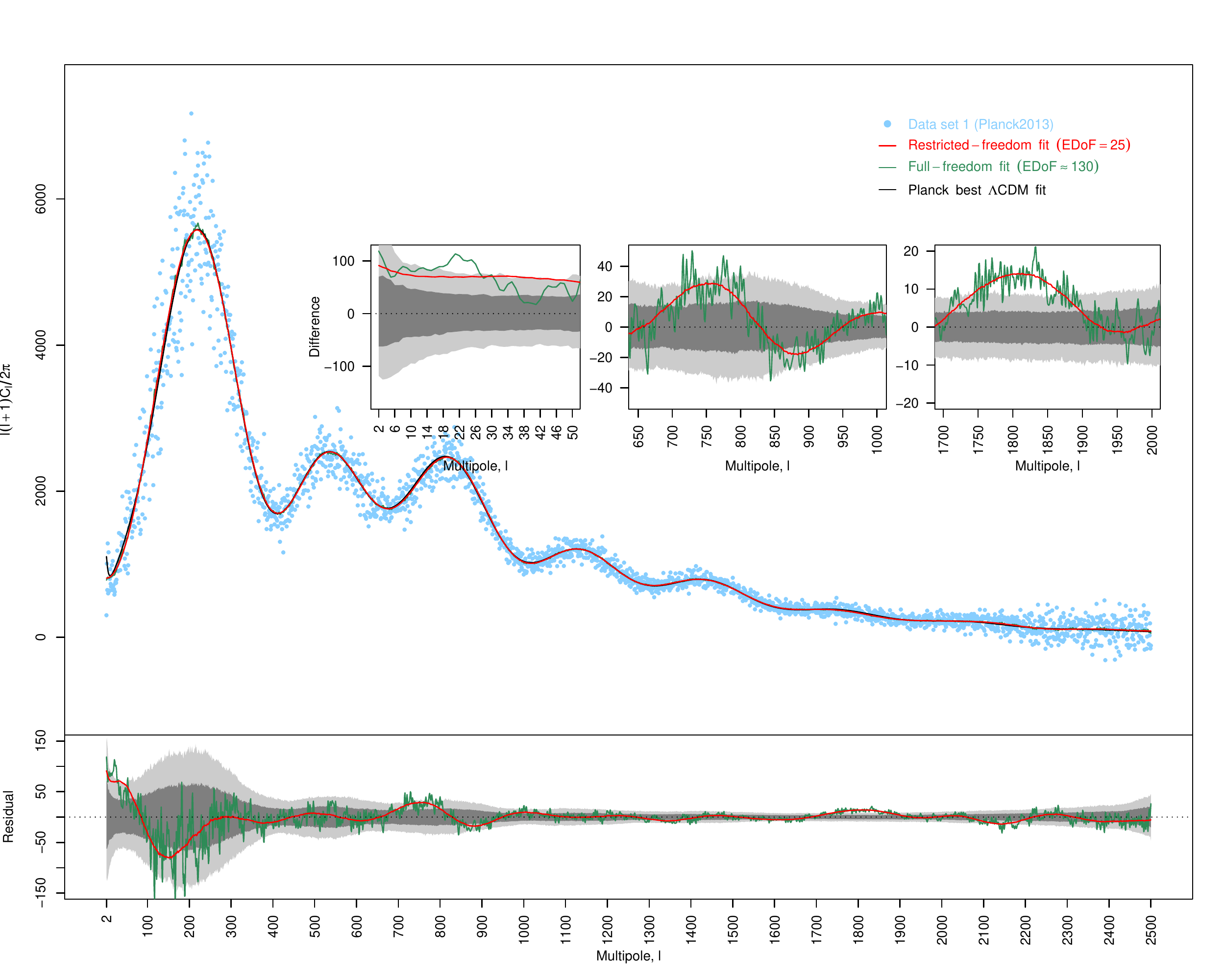}
  \caption{\label{fig:set1plots} (Top panel) The nonparametric fits
for Planck 2013 temperature angular power spectrum data (data set 1).
Green: full-freedom fit (EDoF $\sim 130$), red: restricted-freedom fit
(EDoF $= 25$), black: best fit $\Lambda$CDM model, blue: data set 1,
Planck 2013 angular power spectrum data. The bottom panel illustrates
the residuals of the best fit $\Lambda$CDM model with respect to unbiased nonparametric fits. The $1\sigma$ and $2\sigma$ confidence bands are
plotted in dark and light gray colors. Results show the deviation of
the best fit
$\Lambda$CDM model from data in three multipole ranges (highlighted in
top panel inset plots). These deviations are at low multipoles $18
< \ell < 26$ with greater than $2\sigma$ significance, $\ell\sim750$
at $1$ to $\sim 2 \sigma$ significance and $\ell\sim1800$ at greater than
$2\sigma$ significance.}
 \end{figure}

\begin{figure}
   \includegraphics[width=\textwidth]{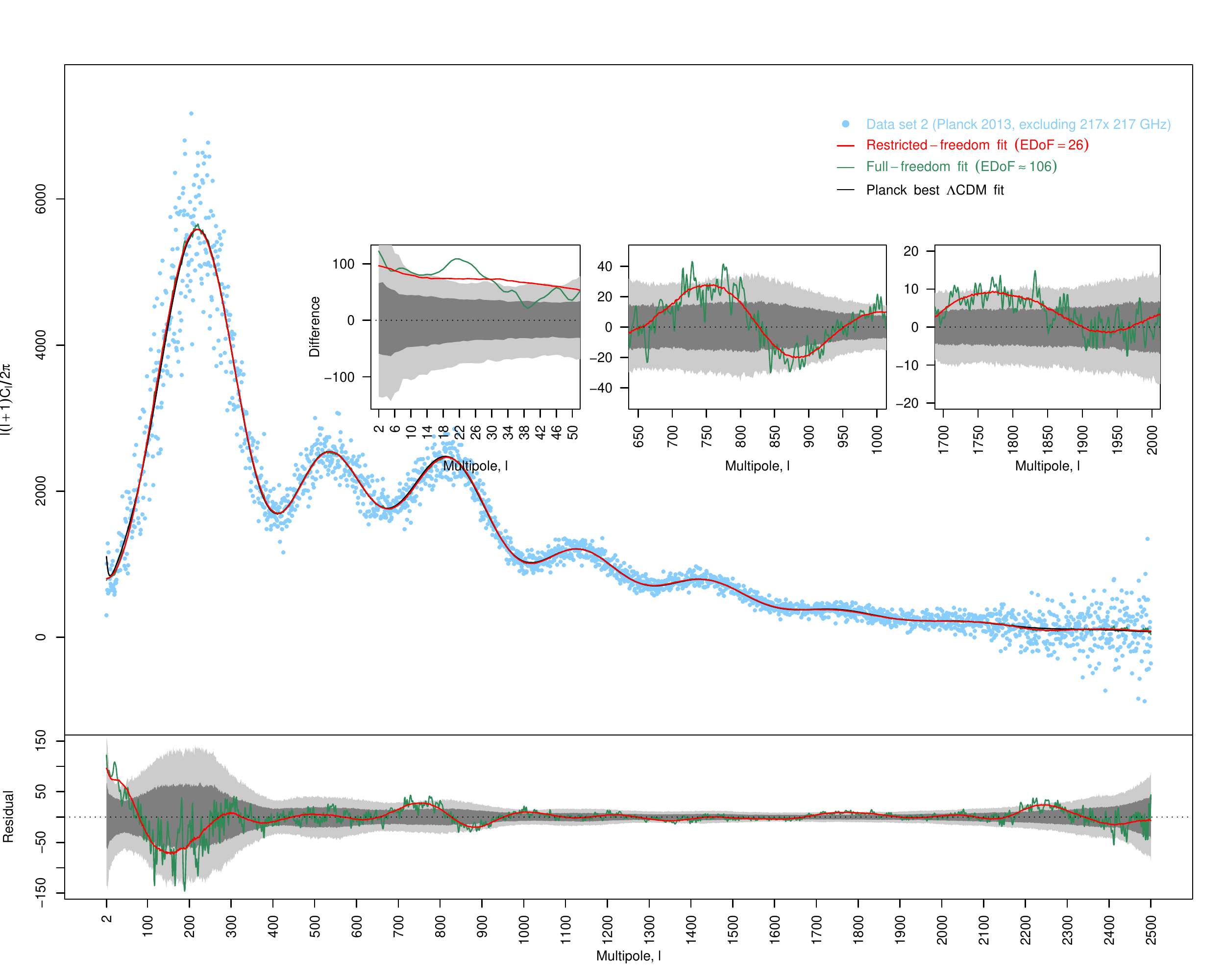}
  \caption{\label{fig:set2plots} (Top panel) The nonparametric fits
for Planck 2013 temperature angular power spectrum data excluding $217 \times 217$ GHz spectrum (data set 2). Green: full-freedom fit (EDoF $\sim
106$), red: restricted-freedom fit (EDoF $= 26$), black: best fit
$\Lambda$CDM model, blue: data set 2, Planck angular power spectrum
data (excluding $217 \times 217$ GHz power spectrum). The bottom panel
illustrates the residuals of the best fit $\Lambda$CDM model with respect to unbiased nonparametric fits. The $1\sigma$ and $2\sigma$
confidence bands are plotted in dark and light gray colors
respectively. The residual of unbiased full-freedom fit (green curve)
and unbiased restricted-freedom fit (red curve) show the deviation of
the best fit $\Lambda$CDM model from data in three multipole ranges
(highlighted in top panel inset plots). They show the deviations at
low multipoles $18 < \ell < 26$ with greater than $2\sigma$
significance, $\ell\sim750$ at $1$ to $\sim 2 \sigma$ significance and
$\ell\sim1800$ at $1$ to $\sim 2 \sigma$ significance. The significance of
the feature at $\ell \sim 1800$ is reduced considerably by excluding
the $217 \times 217$ GHz spectrum from the Planck 2013 data.}
 \end{figure}

The full-freedom fit and the restricted-freedom fit of data set 1
(full Planck 2013 data) correspond to EDoF $\sim130$ and EDoF $=25$
respectively. Figure~\ref{fig:set1plots} (top panel) shows the
full-freedom fit and the restricted-freedom fit of data set 1 with
green and red points. Applying the same nonparametric methodology
(described in Subsection~\ref{subsec:nonparametricfit}) on data set 2
(Planck 2013 data excluding $217 \times 217$ GHz) we achieve the full-freedom
fit at EDoF $\sim106$ and the restricted-freedom fit at EDoF $=26$.
The full-freedom fits (Figures~\ref{fig:set1plots} and \ref{fig:set2plots}, green points) turn
out to be a little wiggly most likely due to the noise in the
corresponding data set. Rather, the restricted-freedom fits
(Figures~\ref{fig:set1plots} and \ref{fig:set2plots}, red points) exhibit a smooth angular
power spectrum with $6$ peaks up to $\ell \sim2000$ as expected in the
standard cosmology. We see for both sets of data, the
restricted-freedom fits follow the corresponding full-freedom fits
closely. For comparison we also plot the Planck best fit $\Lambda$CDM
(Figures~\ref{fig:set1plots} and \ref{fig:set2plots}, black points).
We see all three fits by and large follow the same trend except for
small deviations. In particular, the best fit $\Lambda$CDM model shows
an up-turn at $\ell<10$ which we expect to see in the concordance
model of cosmology due to the integrated Sachs-Wolfe effect
\citep{Sachs_Wolfe}. However, neither the restricted-freedom
nonparametric fits nor the full-freedom fits, reveal such an upturn.

\begin{figure}
   \includegraphics[width=1\textwidth]{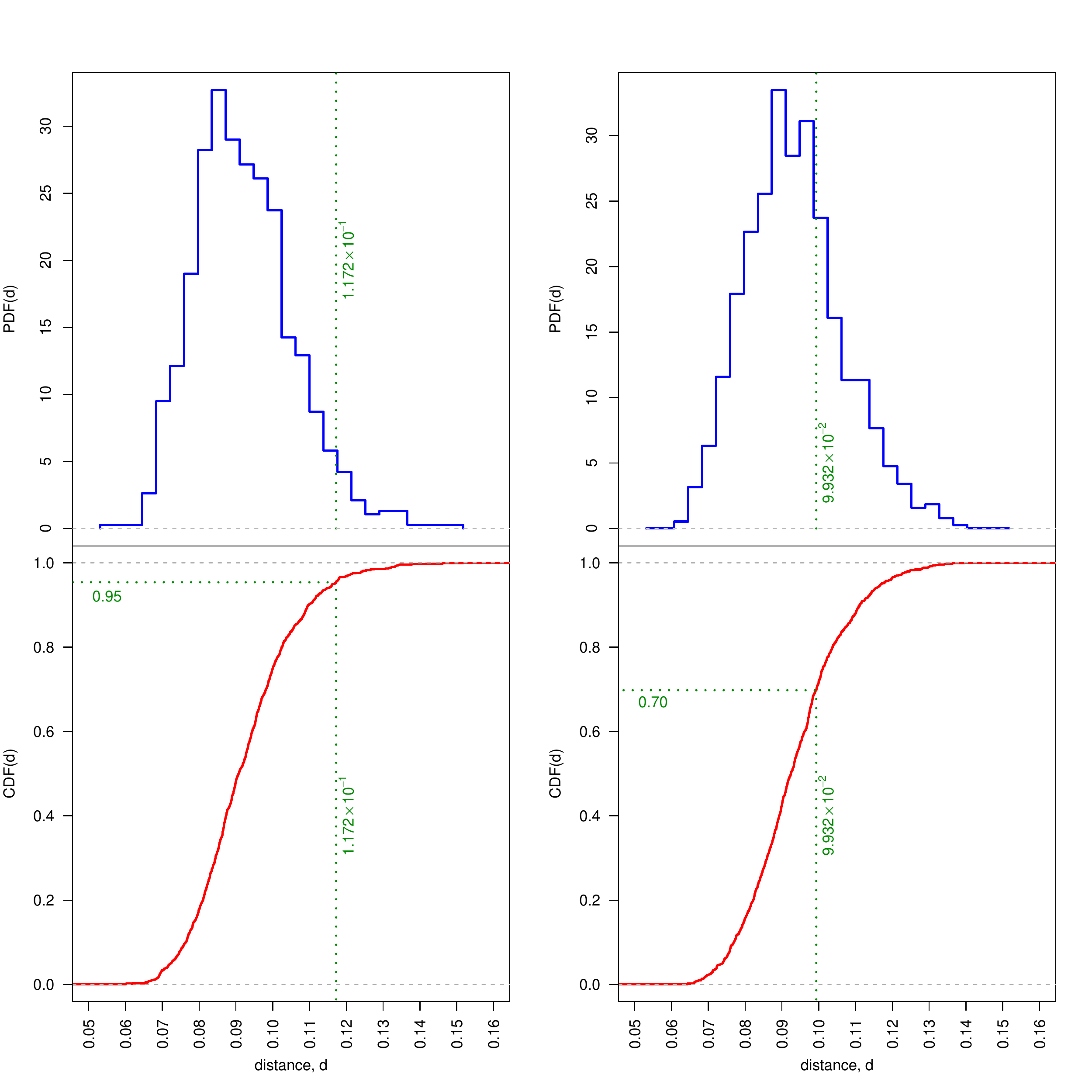}
  \caption{\label{fig:set1pdf} The estimated Probability Density
Functions (PDF) of the distance parameter, $d$, (Equation~\ref{eq:distance}) from
1000 simulations of the Planck 2013 data (based on the best fit
$\Lambda$CDM model) for data set 1 (top-left panel) and data set 2
(top-right panel). The bottom panels shows the empirical Cumulative
Distribution Function (CDF) corresponding to PDF of data set 1
(left-bottom panel) and data set 2 (right-bottom panel). In
comparison and in the case of the actual real data, the Planck 2013
best fit $\Lambda$CDM stands at $1.172 \times 10^{-1}$ distance
(left-bottom panel, green vertical dotted line) from the corresponding
full-freedom fit in the function space. This can be interpreted that
the best fit $\Lambda$CDM model is consistent to the Planck 2013 data
at $95\% (\sim 2\sigma)$ confidence level (left-bottom panel, green
horizontal dotted line). Excluding the $217 \times 217$ GHz spectrum (data
set 2), the Planck best fit $\Lambda$CDM stands in $9.932 \times
10^{-2}$ distance (right-bottom panel, green vertical dotted line)
with respect to the corresponding full-freedom fit in the function
space which make this model to be consistent to the data set 2
(Planck 2013 excluding $217 \times 217$ GHz spectrum) at $70\% (\sim 1\sigma)$
confidence level (right-bottom panel, green horizontal dotted line).}
 \end{figure}

For testing the consistency of the Planck best fit $\Lambda$CDM and
the data we employ the Monte Carlo simulation described in Subsection~\ref{subsec:calibrating}.
The simulations mimic 1000 independent observations of angular power
spectrum for data set 1 and 2 all based on the best fit $\Lambda$CDM
model. Applying the nonparametric method we obtain the full-freedom
fits of the simulated data and the corresponding distances to the
Planck best fit $\Lambda$CDM (as the true model). The estimated
Probability Density Functions (PDF) of the calculated distances are
shown for data set 1 and data set 2, in Figure~\ref{fig:set1pdf},
top-left and top-right panels respectively. Furthermore the
corresponding empirical Cumulative Distribution Functions (CDF) are
plotted in Figure~\ref{fig:set1pdf} (bottom panels). These plots will
provide us a scale that at what distances the best fit $\Lambda$CDM
model and the full-freedom non-parametric fits should stand if the
observed data is a true realization of the concordance model.

Using the Equation~\ref{eq:distance}, the Planck best fit $\Lambda$CDM
stands in function space at the distances $1.172 \times 10^{-1}$ and
$9.932 \times 10^{-2}$ to the full-freedom fits associated with data
set 1 and data set 2 respectively. Locating these numbers in the CDF
diagrams of the data set 1 and 2 (Figure~\ref{fig:set1pdf}, bottom
panels) we find that the corresponding confidence levels of these
distances are $0.95$ and $0.70$ respectively.

Deriving the confidence levels, we also attempt to transfer these
information from the function space to multipole $\ell$ angular power
spectrum space. As described in Subsection~\ref{subsec:bias}, we try
to remove the bias from the nonparametric fits in order to compare
them with the best fit $\Lambda$CDM model. We use distribution of 1000
full-freedom fits at each $\ell$ obtained from the simulated angular
power spectrum data (which are all based on the best fit $\Lambda$CDM
model) and calculate the median values, $\widehat{{\cal
D}}^{med}_{\ell}$. We use these median values at each $\ell$ to derive
residuals through Equation~\ref{eq:residual2}.
Figures~\ref{fig:set1plots} and \ref{fig:set2plots} (bottom panels)
illustrate the residual values of the best fit $\Lambda$CDM model with respect to the (unbiased) full-freedom fit (green curve) using data
set 1 and data set 2 respectively. The median, $\widehat{{\cal
D}}^{med}_{\ell}$ is indicated as the zero line (black dotted line) in
these plots. The $1\sigma$ and $2\sigma$ confidence bands based on the
simulations are shown by dark and light gray colors. We should note
here that since in Figures~\ref{fig:set1plots} and \ref{fig:set2plots}
(bottom panels) we plot the residuals of the best fit $\Lambda$CDM model with respect to the nonparametric fit the zero line represents
the nonparametric fit. From Figure~\ref{fig:set1plots} (bottom
panel) we see that the most prominent deviations between the
nonparametric reconstructions and the best fit $\Lambda$CDM model
seems to be at low multipoles $18 < \ell < 26$ at greater than
$2\sigma$ significance, $\ell\sim750$ at $1$ to $\sim 2 \sigma$ significance
and $\ell\sim1800$ at greater than $2\sigma$ significance. For better
illustration these three regions are replotted in top panel, inset
plots. Figure~\ref{fig:set1plots} (bottom panel) shows the similar
plots for data set 2 (Planck 2013 excluding $217 \times 217$ GHz spectrum).
We see that the feature at $\ell \sim 1800$ becomes substantially less
significant at $1$ to $\sim 2\sigma$ confidence level. Considering all
these, the deviations at low multipoles and $\ell \sim 750$ looks to
be the most prominent feature in the Planck 2013 data.

\section{Conclusion}~\label{sec:conclusions}
In this paper we have calibrated REACT confidence distances to present an algorithm to accurately estimate the consistency of an assumed theoretical cosmological model with CMB data. Using our
nonparametric method we reconstruct the form of the angular power spectrum from the data and compare our results with the theoretically
expected angular power spectrum. We recently performed an analysis of comparing the best fit $\Lambda$CDM model and
Planck data using REACT in~\cite{Aghamousa2014b} where we have calculated the confidence distance of the standard $\Lambda$CDM model from the center of the nonparametric confidence set.

However, in this work in order to have a more accurate quantitative statistical measure of comparison between the nonparmetric results
and the model theoretical expectations we calibrated our consistency measure using many realizations of the data all based on the same
theoretical model. While the actual REACT is very conservative in estimation of the inconsistencies between data sets and models and we
found previously that the best fit $\Lambda$CDM model is at $36\% (\sim 0.5\sigma)$ confidence distance from the center of the nonparametric confidence
set using Planck 2013 temperature data, our calibrated results put the best fit $\Lambda$CDM model at $95\% (\sim 2\sigma)$ confidence distance from the center
of the nonparametric confidence set. In this work we have simulated 1000 Monte Carlo realizations of the Planck 2013 temperature angular
power spectrum data based on the best fit $\Lambda$CDM model and for each realization we derived the confidence distance between the best nonparametric fit and the true model. We
used the distribution of these confidence distances from the simulations to calibrate REACT in this particular problem and we compared the derived distribution with the results from the actual data.
Our results indicate that the distance between the best fit $\Lambda$CDM
model from the center of the nonparametric confidence set in the case of the real Planck
2013 temperature data is larger than the similar distances we derived in simulations in $95\%$ of
the cases. Our results seems to be in a very close agreement with some other analysis
discussing the same subject using different methodologies ~\cite{Hazra2014a, Hazra2014b, Larson2014}. Excluding the 217 GHz data
from our analysis, the best fit $\Lambda$CDM model shifts to $70\% (\sim 1\sigma)$ confidence distance from the
center of nonparametric confidence set resulting to better consistency of the standard model
with Planck data. While the $217 \times 217$ spectrum is one of the most precise CMB measurements
of the Planck, in~\cite{Planck2013XXII} it has been mentioned that there has been a small systematic
in this frequency channel (apparently due to incomplete removal of 4 K cooler lines). This
probably can be a reason behind the significant inconsistency we derived between the standard
$\Lambda$CDM model and full Planck 2013 data. Excluding the $217 \times 217$ spectrum from our
analysis and getting significantly better consistency between Planck 2013 and the
concordance model seems to support this argument. However, repetition of the analysis using
Planck 2015 can make things much more clear.

Using this algorithm we can also highlight the angular scales where
the assumed theoretical model deviates from the data (with considerable statistical significance) using large number of simulations
and after some bias control. We should recall that REACT and consequently our calibrated REACT in this analysis is a biased estimator and while this would not
affect our consistency test, to look for angular scales deviating from the concordance model we should perform a bias control. We can perform this bias control using extensive
simulations and subsequently correct/modify the estimated forms of the angular power spectrum.
Applying our approach on Planck temperature 2013 data and assuming the best fit spatially flat $\Lambda$CDM model, after performing the bias control
and error-estimation our results indicate that the most prominent features in the Planck 2013 data deviating from the best fit $\Lambda$CDM model seems to be at low multipoles
$ 18 < \ell < 26$ with greater than $2\sigma$ significance, $\ell\sim750$ at $1$ to $\sim2\sigma$
significance and $\ell\sim1800$ with greater than $2\sigma$ significance. Excluding
the 217 GHz data the feature at $\ell\sim1800$ becomes substantially less
significant at $1$ to $\sim2\sigma$. While the feature at $\ell\sim1800$ seems to
be due to some systematics in the data, the features at $\ell\sim750$
looks to be the most prominent feature in the Planck 2013 data.
Similar results have been reported earlier using alternative
methods~\cite{Hazra2014PPS}.

To summarize, in this paper we propose an approach based on calibration of REACT
 confidence distances where we can test the consistency of a particular theoretical model
with CMB observations and also look for the multipole scales where there are significant
deviations. We have applied the method on Planck 2013 data in order to show its strength and
simplicity and it can be trivially used against forthcoming data including Planck 2015.

\section*{Acknowledgments}

We would like to thank Stephen Appleby, Mihir Arjunwadkar, Dhiraj
Hazra and Tarun Souradeep for useful discussions. 
The authors wish to acknowledge support from the Korea Ministry of Education, Science and Technology, 
Gyeongsangbuk-Do and Pohang City for Independent Junior Research Groups at the Asia Pacific Center for Theoretical Physics.
AS would acknowledge the support of the National Research Foundation of Korea
(NRF-2013R1A1A2013795). The authors acknowledge the use of Planck data
and likelihood from Planck Legacy Archive (PLA). Throughout this work
we use \texttt{R} statistical computing environment \cite{R} in both
computational and plotting tasks.

\bibliographystyle{JHEP}

\begin{thebibliography}{99}

\bibitem{Bennett2013}
C.~L. Bennett et al. {\it {Nine-year Wilkinson Microwave Anisotropy Probe (WMAP)
  Observations: Final Maps and Results}},  {\em Astrophys. J. Supplement} {\bf
  208} (Oct., 2013) 20, [\href{http://xxx.lanl.gov/abs/1212.5225}{{\tt
  arXiv:1212.5225}}].

\bibitem{Hinshaw2013}
G.~Hinshaw et al. {\it {Nine-year Wilkinson Microwave Anisotropy Probe (WMAP)
  Observations: Cosmological Parameter Results}},  {\em Astrophys. J.
  Supplement} {\bf 208} (Oct., 2013) 19,
  [\href{http://xxx.lanl.gov/abs/1212.5226}{{\tt arXiv:1212.5226}}].

\bibitem{Planck2013I}
P. A. R. Ade et al. {\it Planck 2013 results. I. overview of products and scientific results},
  {\em A\&A} {\bf 571} (2014) A1.

\bibitem{Seievers2013ACT}
J.~L. Sievers et al. {\it {The Atacama Cosmology
  Telescope: cosmological parameters from three seasons of data}},  {\em J.
  Cosmology Astropart. Phys.} {\bf 10} (Oct., 2013) 60,
  [\href{http://xxx.lanl.gov/abs/1301.0824}{{\tt arXiv:1301.0824}}].

\bibitem{Story2013SPT}
K.~T. Story et al. {\it {A Measurement of the Cosmic Microwave Background Damping
  Tail from the 2500-Square-Degree SPT-SZ Survey}},  {\em Astrophys. J.} {\bf
  779} (Dec., 2013) 86, [\href{http://xxx.lanl.gov/abs/1210.7231}{{\tt
  arXiv:1210.7231}}].

\bibitem{Planck2013XV}
P. A. R. Ade et al. {\it Planck 2013 results. XV. cmb power spectra and
  likelihood},  {\em A\&A} {\bf 571} (2014) A15.

\bibitem{Planck2013XVI}
P. A. R. Ade et al. {\it Planck 2013 results. XVI. cosmological parameters},  {\em A\&A} {\bf 571} (2014) A16.

\bibitem{Planck2013XXXI}
P. A. R. Ade et al. {\it Planck 2013 results. XXXI. consistency of the planck data},  {\em A\&A} {\bf
  571} (2014) A31.

\bibitem{Kovacs2013}
A.~{Kov{\'a}cs}, J.~{Carron}, and I.~{Szapudi}, {\it {On the coherence of WMAP
  and Planck temperature maps}},  {\em Mon.Not.Roy.Astron.Soc} {\bf 436} (Dec.,
  2013) 1422--1429, [\href{http://xxx.lanl.gov/abs/1307.1111}{{\tt
  arXiv:1307.1111}}].

\bibitem{Hazra2014b}
D.~K. {Hazra} and A.~{Shafieloo}, {\it {Test of consistency between Planck and
  WMAP}},  {\em Phys. Rev. D} {\bf 89} (Feb., 2014) 043004,
  [\href{http://xxx.lanl.gov/abs/1308.2911}{{\tt arXiv:1308.2911}}].

\bibitem{Hazra2014a}
D.~K. {Hazra} and A.~{Shafieloo}, {\it {Confronting the concordance model of
  cosmology with Planck data}},  {\em Journal of Cosmology and Astroparticle Physics} {\bf 1}
  (Jan., 2014) 43, [\href{http://xxx.lanl.gov/abs/1401.0595}{{\tt
  arXiv:1401.0595}}].

\bibitem{Larson2014}
D.~{Larson}, J.~L. {Weiland}, G.~{Hinshaw}, and C.~L. {Bennett}, {\it
  {Comparing Planck and WMAP: Maps, Spectra, and Parameters}},  {\em ArXiv
  e-prints} (Sept., 2014) [\href{http://xxx.lanl.gov/abs/1409.7718}{{\tt
  arXiv:1409.7718}}].

\bibitem{Aghamousa2012}
A.~Aghamousa, M.~Arjunwadkar, and T.~Souradeep, {\it Evolution of the cosmic
  microwave background power spectrum across wilkinson microwave anisotropy
  probe data releases: A nonparametric analysis},  {\em Astrophys. J.} {\bf
  745} (2012), no.~2 114.

\bibitem{Aghamousa2014b}
A.~Aghamousa, A.~Shafieloo, M.~Arjunwadkar, and T.~Souradeep, {\it Unveiling
  acoustic physics of the cmb using nonparametric estimation of the temperature
  angular power spectrum for planck},  {\em Journal of Cosmology and
  Astroparticle Physics} {\bf 2015} (2015), no.~02 007.

\bibitem{Planck2013XXII}
P. A. R. Ade et al. {\it Planck 2013 results. XXII. constraints on inflation},  {\em A\&A} {\bf 571} (2014) A22.

\bibitem{Spergel2013}
D.~{Spergel}, R.~{Flauger}, and R.~{Hlozek}, {\it {Planck Data Reconsidered}},
  {\em ArXiv e-prints} (Dec., 2013)
  [\href{http://xxx.lanl.gov/abs/1312.3313}{{\tt arXiv:1312.3313}}].

\bibitem{Beran1996}
R.~Beran, {\it Confidence sets centred at $c_p$-estimators},  {\em Ann.\ Inst.\
  Statist.\ Math.} {\bf 48} (1996) 1--15.

\bibitem{BD1998}
R.~Beran and L.~D\"umbgen, {\it Modulation of estimators and confidence sets},
  {\em Ann.\ Statist.} {\bf 26} (1998), no.~5 1826--1856.

\bibitem{Beran2000}
R.~Beran, {\it React scatterplot smoothers: Superefficiency through basis
  economy},  {\em J.\ Amer.\ Statist.\ Assoc.} {\bf 95} (2000), no.~449
  155--171.

\bibitem{Beran2000b}
R.~Beran, {\it React trend estimation in correlated noise},  in {\em
  Asymptotics in statistics and probability: papers in honor of George Gregory
  Roussas} (M.~L. Puri, ed.), pp.~1--16.
\newblock VSP International Science Publishers, 2000.

\bibitem{GMN+2004}
C.~R. Genovese, C.~J. Miller, R.~C. Nichol, M.~Arjunwadkar, and L.~Wasserman,
  {\it Nonparametric inference for the cosmic microwave background},  {\em
  Statist. Sci.} {\bf 19} (2004), no.~2 308--321.

\bibitem{BSM+2007}
B.~Bryan, J.~Schneider, C.~J. Miller, R.~C. Nichol, C.~R. Genovese, and
  L.~Wasserman, {\it Mapping the cosmological confidence ball surface},  {\em
  Astrophys. J.} {\bf 665} (August, 2007) 25--41.

\bibitem{Aghamousa2014}
A.~Aghamousa, M.~Arjunwadkar, and T.~Souradeep, {\it Model-independent
  forecasts of cmb angular power spectra for the planck mission},  {\em Phys.
  Rev. D} {\bf 89} (Jan, 2014) 023509.

\bibitem{Stein1981}
C.~M. Stein, {\it Estimation of the mean of a multivariate normal
  distribution},  {\em The Annals of Statistics} {\bf 9} (1981), no.~6
  1135--1151.

\bibitem{Sachs_Wolfe}
R.~K. {Sachs} and A.~M. {Wolfe}, {\it {Perturbations of a Cosmological Model
  and Angular Variations of the Microwave Background}},  {\em Astrophys. J.}
  {\bf 147} (Jan., 1967) 73.

\bibitem{Hazra2014PPS}
D.~K. {Hazra}, A.~{Shafieloo}, and T.~{Souradeep}, {\it {Primordial power
  spectrum from Planck}},  {\em Journal of Cosmology and Astroparticle Physics} {\bf 11} (Nov.,
  2014) 11, [\href{http://xxx.lanl.gov/abs/1406.4827}{{\tt arXiv:1406.4827}}].

\bibitem{R}
{R Core Team}, {\em R: A Language and Environment for Statistical Computing}.
\newblock R Foundation for Statistical Computing, Vienna, Austria, 2013.

\end{thebibliography}

\end{document}